\documentclass[aps,prc,superscriptaddress,showpacs,twocolumn,floatfix]{revtex4}
\usepackage{graphicx}
\usepackage[dvips]{color}
\usepackage{epsfig}
\begin{document}

\title{Demonstration of a novel technique to measure two-photon exchange
effects in elastic $e^\pm p$ scattering}

\newcommand*{\FIU}{Florida International University, Miami, Florida
  33199, USA} 
\newcommand*{\FIUindex}{12} \affiliation{\FIU}
\newcommand*{\ODU}{Old Dominion University, Norfolk, Virginia 23529,
  USA} 
\newcommand*{\ODUindex}{29} \affiliation{\ODU}
\newcommand*{\ANL}{Argonne National Laboratory, Argonne, Illinois
  60439, USA} 
\newcommand*{\ANLindex}{1} \affiliation{\ANL}
\newcommand*{\UTFSM}{Universidad T\'{e}cnica Federico Santa Mar\'{i}a,
  Casilla 110-V Valpara\'{i}so, Chile} 
\newcommand*{\UTFSMindex}{36}\affiliation{\UTFSM} 
\newcommand*{\UCONN}{University of Connecticut, Storrs, Connecticut
  06269, USA}   
\newcommand*{\UCONNindex}{9}\affiliation{\UCONN} 
\newcommand*{\JLAB}{Thomas Jefferson National Accelerator Facility,
  Newport News, Virginia 23606, USA}  
\newcommand*{\JLABindex}{35} \affiliation{\JLAB}
\newcommand*{\ASU}{Arizona State University, Tempe, Arizona 85287, USA} 
\newcommand*{\ASUindex}{2} \affiliation{\ASU}
\newcommand*{\CSUDH}{California State University, Dominguez Hills,
  Carson, California 90747, USA} 
\newcommand*{\CSUDHindex}{3}\affiliation{\CSUDH} 
\newcommand*{\CANISIUS}{Canisius College, Buffalo, New York 14208,
  USA} 
\newcommand*{\CANISIUSindex}{4} \affiliation{\CANISIUS}
\newcommand*{\CMU}{Carnegie Mellon University, Pittsburgh Pennsylvania
  15213, USA} 
\newcommand*{\CMUindex}{5} \affiliation{\CMU}
\newcommand*{\CUA}{Catholic University of America, Washington,
  D.C. 20064, USA}  
\newcommand*{\CUAindex}{6} \affiliation{\CUA}
\newcommand*{\SACLAY}{CEA, Centre de Saclay, Irfu/Service de Physique
  Nucl\'eaire, 91191 Gif-sur-Yvette, France}
\newcommand*{\SACLAYindex}{7} \affiliation{\SACLAY}
\newcommand*{\CNU}{Christopher Newport University, Newport News,
  Virginia 23606, USA} 
\newcommand*{\CNUindex}{8} \affiliation{\CNU}
\newcommand*{\EDINBURGH}{Edinburgh University, Edinburgh EH9 3JZ,
  United Kingdom} 
\newcommand*{\EDINBURGHindex}{10}\affiliation{\EDINBURGH} 
\newcommand*{\FU}{Fairfield University, Fairfield, Connecticut 06824,
  USA}  
\newcommand*{\FUindex}{11} \affiliation{\FU}
\newcommand*{\FSU}{Florida State University, Tallahassee, Florida
  32306, USA} 
\newcommand*{\FSUindex}{13} \affiliation{\FSU}
\newcommand*{\GWUI}{The George Washington University, Washington, DC
  20052, USA} 
\newcommand*{\GWUIindex}{14} \affiliation{\GWUI}
\newcommand*{\ISU}{Idaho State University, Pocatello, Idaho 83209, USA}
\newcommand*{\ISUindex}{15} \affiliation{\ISU}
\newcommand*{\INFNFE}{INFN, Sezione di Ferrara, 44100 Ferrara, Italy}
\newcommand*{\INFNFEindex}{16} \affiliation{\INFNFE}
\newcommand*{\INFNFR}{INFN, Laboratori Nazionali di Frascati, 00044
  Frascati, Italy} 
\newcommand*{\INFNFRindex}{17}\affiliation{\INFNFR} 
\newcommand*{\INFNGE}{INFN, Sezione di Genova, 16146 Genova, Italy} 
\newcommand*{\INFNGEindex}{18}\affiliation{\INFNGE} 
\newcommand*{\INFNRO}{INFN, Sezione di Roma Tor Vergata, 00133 Rome,
  Italy} 
\newcommand*{\INFNROindex}{19}\affiliation{\INFNRO} 
\newcommand*{\ORSAY}{Institut de Physique Nucl\'eaire ORSAY, Orsay,
  France} 
\newcommand*{\ORSAYindex}{20}\affiliation{\ORSAY} 
\newcommand*{\ITEP}{Institute of Theoretical and Experimental Physics,
  Moscow, 117259, Russia} 
\newcommand*{\ITEPindex}{21} \affiliation{\ITEP}
\newcommand*{\JMU}{James Madison University, Harrisonburg, Virginia
  22807, USA} 
\newcommand*{\JMUindex}{22} \affiliation{\JMU}
\newcommand*{\KNU}{Kyungpook National University, Daegu 702-701,
  Republic of Korea} 
\newcommand*{\KNUindex}{23} \affiliation{\KNU}
\newcommand*{\LPSC}{LPSC, Universite Joseph Fourier, CNRS/IN2P3, INPG,
  Grenoble, France} 
\newcommand*{\LPSCindex}{24}\affiliation{\LPSC} 
\newcommand*{\UNH}{University of New Hampshire, Durham, New Hampshire
  03824, USA} 
\newcommand*{\UNHindex}{25}\affiliation{\UNH} 
\newcommand*{\NSU}{Norfolk State University, Norfolk, Virginia 23504,
  USA} 
\newcommand*{\NSUindex}{26}\affiliation{\NSU} 
\newcommand*{\OHIOU}{Ohio University, Athens, Ohio 45701, USA} 
\newcommand*{\OHIOUindex}{27}\affiliation{\OHIOU}
\newcommand*{\RPI}{Rensselaer Polytechnic Institute, Troy, New York
  12180, USA} 
\newcommand*{\RPIindex}{29}\affiliation{\RPI}
\newcommand*{\URICH}{University of Richmond, Richmond, Virginia 23173,
  USA}
\newcommand*{\URICHindex}{30}\affiliation{\URICH}
\newcommand*{\ROMAII}{Universita' di Roma Tor Vergata, 00133 Rome
  Italy} 
\newcommand*{\ROMAIIindex}{31}\affiliation{\ROMAII}
\newcommand*{\MSU}{Skobeltsyn Nuclear Physics Institute, 119899
  Moscow, Russia} 
\newcommand*{\MSUindex}{32}\affiliation{\MSU}
\newcommand*{\SCAROLINA}{University of South Carolina, Columbia, South
  Carolina 29208, USA} 
\newcommand*{\SCAROLINAindex}{33}\affiliation{\SCAROLINA} 
\newcommand*{\GLASGOW}{University of Glasgow, Glasgow G12 8QQ, United
  Kingdom} 
\newcommand*{\GLASGOWindex}{36}\affiliation{\GLASGOW} 
\newcommand*{\VIRGINIA}{University of Virginia, Charlottesville,
  Virginia 22901, USA} 
\newcommand*{\VIRGINIAindex}{37}\affiliation{\VIRGINIA} 
\newcommand*{\WM}{College of William and Mary, Williamsburg, Virginia
  23187, USA} 
\newcommand*{\WMindex}{38}\affiliation{\WM} 
\newcommand*{\YEREVAN}{Yerevan Physics Institute, 375036 Yerevan,
  Armenia} 
\newcommand*{\YEREVANindex}{39}\affiliation{\YEREVAN}

\newcommand*{\NOWCNU}{Christopher Newport University, Newport News, Virginia 23606}
\newcommand*{\NOWLANL}{Los Alamos National Laboratory, Los Alamos, NM 87544 USA}
\newcommand*{\NOWMSU}{Skobeltsyn Nuclear Physics Institute, 119899 Moscow, Russia}
\newcommand*{\NOWIU}{Indiana University, Bloomington, IN 47405}
\newcommand*{\NOWORSAY}{Institut de Physique Nucl\'eaire ORSAY, Orsay, France}
\newcommand*{\NOWROMAII}{Universita' di Roma Tor Vergata, 00133 Rome Italy}

\author{M. Moteabbed}
\affiliation{\FIU}
\author{M. Niroula}
\affiliation{\ODU}
\author{B.A. Raue}
\affiliation{\FIU}
\author{L.B. Weinstein}
\affiliation{\ODU}
\author{D. Adikaram}
\affiliation{\ODU}
\author{J. Arrington}
\affiliation{\ANL} 
\author{W.K. Brooks}
\affiliation{\UTFSM}
\author{J. Lachniet}
\affiliation{\ODU}
\author{Dipak Rimal}
\affiliation{\FIU}
\author{M. Ungaro}
\affiliation{\UCONN}
\affiliation{\JLAB}
% ...

\collaboration{The CLAS Collaboration}
\noaffiliation

\date{\today}

\begin{abstract}
\begin{description}

\item[Background:]  The discrepancy between proton electromagnetic form
factors extracted using unpolarized and polarized scattering data is believed
to be a consequence of two-photon exchange (TPE) effects.  However, the 
calculations of TPE corrections have significant model dependence, and 
there is limited direct experimental evidence for such corrections.

\item[Purpose:] The TPE contributions depend on the sign of the lepton charge
in $e^\pm p$ scattering, but the luminosities of secondary positron beams 
limited past measurement at large scattering angle where the TPE effects are
believe to be most significant. We present the results of a new experimental
technique for making direct $e^\pm p$ comparisons, which has the potential to
make precise measurements over a broad range in $Q^2$ and scattering angles.

\item[Methods:] We use the Jefferson Lab electron beam and the Hall
B photon tagger to generate a clean but untagged photon beam.  The photon beam
impinges on a converter foil to generate a mixed beam of electrons, positrons,
and photons.  A chicane is used to separate and recombine the electron and
positron beams while the photon beam is stopped by a photon blocker.  This
provides a combined electron and positron beam, with energies from 0.5 to 3.2
GeV, which impinges on a liquid hydrogen target.  The large acceptance CLAS
detector is used to identify and reconstruct elastic scattering events,
determining both the initial lepton energy and the sign of the scattered
lepton.

\item[Results:] The data were collected in two days with a primary electron
beam energy of only 3.3 GeV, limiting the data from this run to smaller values
of $Q^2$ and scattering angle.  Nonetheless, this measurement yields a data
sample for $e^\pm p$ with statistics comparable to those of the best previous 
measurements. We have shown that
we can cleanly identify elastic scattering events and correct for the
difference in acceptance for electron and positron scattering.  Because we ran
with only one polarity for the chicane, we are unable to study the difference
between the incoming electron and positron beams. This systematic effect leads 
to the largest uncertainty in the final ratio of positron to electron scattering:
$R=1.027\pm0.005\pm0.05$ for $\langle Q^2\rangle=0.206$ GeV$^2$ and
$0.830\leq \epsilon\leq 0.943$.

\item[Conclusions:] We have demonstrated that the tertiary $e^\pm$ beam
generated using this novel technique provides the opportunity for dramatically
improved comparisons of $e^\pm p$ scattering, covering a significant range in
both $Q^2$ and scattering angle.  Combining data with different chicane
polarities will allow for detailed studies of the difference between the
incoming $e^+$ and $e^-$ beams.
\end{description}
\end{abstract}


\maketitle

\section{Introduction}
\label{sec-intro}

Electron scattering is one of the most powerful tools available for
measurements involving the quark structure of nucleons and nuclei.  The
dominant one-photon exchange (OPE) mechanism is well understood, and the
relatively weak electromagnetic coupling means that the scattering
uniformly probes the matter within even a dense nucleus.  This weak
coupling also implies small higher-order corrections to the cross section
related to two-photon exchange (TPE), which are suppressed by an additional
power of the fine structure constant $\alpha \approx 1/137$.  Thus,
electron scattering is the primary probe of the structure of stable
hadrons, and in particular of the elastic electromagnetic form factors of
the proton~\cite{arrington07a, perdrisat07, arrington11a}.

There is renewed interest in two-photon exchange contributions due to
new polarization measurements of the proton electromagnetic form
factors, $G_E(Q^2)$ and $G_M(Q^2)$.  High-$Q^2$ measurements using
recoil polarization techniques to extract the ratio $\mu_p
G_E/G_M$~\cite{punjabi05, puckett10, puckett11} indicated a
significant discrepancy~\cite{arrington03a} with extractions based on
the Rosenbluth separation technique~\cite{walker94, andivahis94,
  christy04, qattan05}.  This led to the concern that TPE corrections
to the cross section may be more important than previously
thought~\cite{guichon03, blunden03, chen04}, with implications for not
only the form factors, but also other precision measurements using
electron scattering~\cite{arrington04a, afanasev05, arrington07b,
  blunden09, sibirtsev10, blunden11}.

Theoretical investigations suggest that the TPE contributions may be
sufficient to resolve the discrepancy~\cite{arrington07c, carlson07,
arrington11b}.  For the most part, the calculations indicate that the TPE
effects are small, but have a significant angle dependence, increasing in
magnitude at larger scattering angles.  Because the charge form factor,
$G_E(Q^2)$, is related to the angular dependence of the elastic cross section,
the impact of small TPE corrections can be very large if the angular
dependence associated with $G_E$ becomes small, e.g. at large $Q^2$ values. 
Thus, the charge form factor is a special case with exceptional sensitivity to
TPE corrections. Nonetheless, other high-precision measurements may still need
to evaluate these small contributions, and there are essentially no direct 
measurements of TPE that can be used to validate these calculations.

There is clear evidence for TPE contributions in other processes and other
observables~\cite{offermann86, wells01, maas05, androic11, abrahamyan12},
but little direct evidence for TPE contributions in the
unpolarized elastic electron--proton cross section.  The cleanest and most
direct way to study TPE contributions to the cross section is through the
comparison of electron and positron scattering~\cite{arrington04b,
arrington09b}.  The interference between the single-photon exchange and the
TPE diagrams yields the largest TPE contribution to the cross section, and its
sign depends on the sign of the lepton charge.  Most other radiative
corrections are identical for electron and positron scattering, with the only
other charge-dependent contribution being from the interference between lepton
and proton bremsstrahlung, which is relatively small at low $Q^2$ where the
proton momentum is small.

The main difficulty in measuring TPE contributions in fixed-target
experiments is that the low luminosities of the secondary positron beams
have historically limited measurements to regions where the cross section
is large: low $Q^2$ and/or very forward angle scattering.  The TPE
contributions needed to explain the form factor discrepancy are
relatively small, and become important at larger $Q^2$ and scattering
angles.  Thus, a significant increase in the luminosity is required to make
meaningful measurements in the kinematic region of interest.

We present here the results from an experiment that used a novel technique
to make a simultaneous measurement of positron--proton and electron--proton
elastic scattering.  While the data from this brief run are limited to low
$Q^2$ and small scattering angles, the experiment provides statistics
comparable to the best previous measurements. It also demonstrates the
possibility to cover a large range of $Q^2$ and scattering angles with the
precision and accuracy necessary to determine whether TPE corrections can
explain the observed form factor discrepancy.  Such data can also constrain
calculations of the corrections at low-to-moderate $Q^2$ values, allowing
validation of the calculations that may be needed to evaluate potential TPE
impacts beyond elastic scattering.

\section{Two-Photon Exchange}
\label{sec-tpe}

\begin{figure}[htb]  
\includegraphics[width=0.45\textwidth,clip=true]{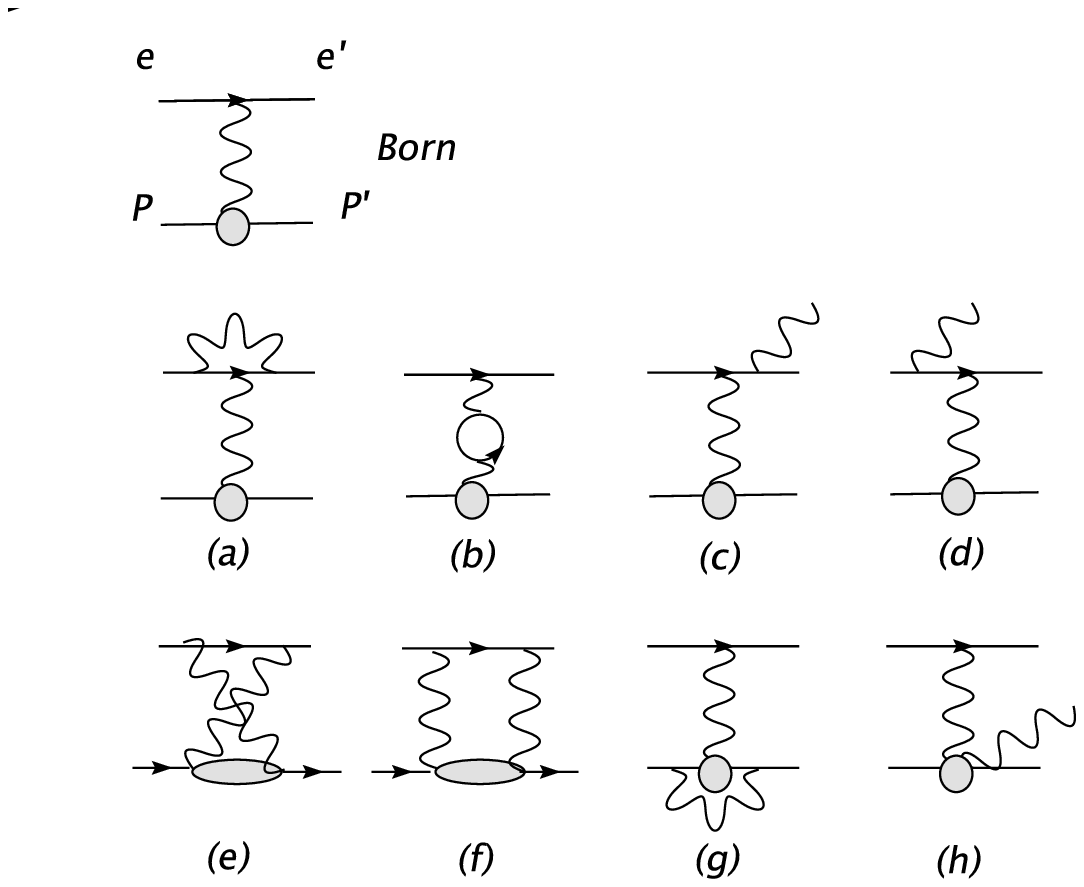} 
\begin{center} 
\caption{\label{fig:diagrams} 
Feynman diagrams for the elastic lepton--proton scattering, including 
the 1st-order QED radiative corrections. Diagram (a) shows the
electron vertex renormalization term, (b) shows the photon propagator
renormalization term, (c) and (d) show the
electron bremsstrahlung terms, (g) shows the proton vertex
renormalization term, (h) shows the proton
bremsstrahlung term, and (e) and (f) show the
two-photon exchange terms, where the intermediate state can be an unexcited
proton, a baryon resonance or a continuum of hadrons.} 
\end{center} 
\end{figure} 

Figure~\ref{fig:diagrams} shows the Born contribution and higher order QED
corrections to lepton--proton elastic scattering. The TPE contribution
(diagrams (e) and (f)) is difficult to calculate because the intermediate
hadronic state must be integrated over all baryonic resonance and continuum
states that can be excited by the virtual photon.  Therefore, TPE is
typically neglected in calculating radiative corrections~\cite{mo69,
tsai71, ent01}, with the exception of the contribution needed to cancel
infrared divergences in bremsstrahlung terms.

A direct measurement of the TPE correction can be
achieved experimentally in the ratio of the positron-proton to
electron-proton elastic cross sections.  Neglecting bremsstrahlung terms,
the Born term and first order corrections from Fig.~\ref{fig:diagrams}
yield a total amplitude for $ep \to ep$ scattering of
\begin{eqnarray}\label{eq:amp}
A_{ep \to ep} = q_e q_p [A_{1\gamma} + q_e^2 A_{e.vertex} + q_p^2 A_{p.vertex} \nonumber \\
+ q_e^2 A_{loop} + q_e q_p A_{2\gamma}],
\end{eqnarray}
where $q_e$ and $q_p$ are the lepton and proton charges and the amplitudes
$A_{1\gamma}, A_{e.vertex}, A_{p.vertex}, A_{e.loop}$ and $A_{2\gamma}$
respectively describe one-photon exchange, electron and proton vertex
corrections [Figs.~\ref{fig:diagrams}(a) and~\ref{fig:diagrams}(g)], loop
corrections [\ref{fig:diagrams}(b)], and two-photon exchange 
[\ref{fig:diagrams}(e) and~\ref{fig:diagrams}(f)]. Squaring the above
amplitude and keeping only the corrections up to order $\alpha$, we have
\begin{eqnarray}\label{eq:sqamp}
|A_{ep\to ep}|^2 \approx e^4 [ ~ A_{1\gamma}^2 
+ 2 e^2 A_{1\gamma} \Re(A_{loop+vertex})  \nonumber \\
+ 2 q_e q_p A_{1\gamma} \Re(A_{2\gamma}) ~],
\end{eqnarray}
where we have simplified the expression by replacing $q_e^2$ and
$q_p^2$ with $e^2$ and taken $A_{loop+vertex}$ to be the sum of the
1st order corrections where the lepton (and proton) charges appear in
even powers, and thus are identical for electron and positron
scattering.  Note that because $A_{1\gamma}$ is real and large
compared to the other terms in Eq.~\ref{eq:amp}, the contribution from
the imaginary part of $A_{2\gamma}$ has a negligible contribution to
the squared amplitude, and it is common to include only the real part
of the TPE amplitude.

Experientally, one cannot always separate true elastic scattering from
events with a radiated photon in the final state.  A cut on the
missing energy or the invariant mass is often used to exclude events
with a high-energy photon in the final state from the case with low
energy reactions.  The interference between electron and proton
bremsstrahlung yields another contribution that changes sign with the
lepton charge, yielding a final cross section that is proportional to
\begin{eqnarray}\label{eq:sqamptot}
  |A_{ep\to ep}|^2 = e^4 \{A_{1\gamma}^2 + 2 e^2 C_{even} \nonumber \\
  + 2 q_e q_p [ A_{1\gamma} \Re(A_{2\gamma}) + \Re(A_{e.br.}^*
  A_{p.br.})] \},
\end{eqnarray}
where $C_{even}$ is the sum of the charge-even part of the radiative
contributions, including both the loop and vertex diagrams and the
charge-even contributions from the electron and proton bremsstrahlung
diagrams [Figs.~\ref{fig:diagrams}(c), ~\ref{fig:diagrams}(d),
and~\ref{fig:diagrams}(h)].  There is no interference between the Born
term and the bremsstrahlung terms because they have different final
states. The only portion of the bremsstrahlung term that is not
charge-even is the interference between $A_{e.br.}$ and $A_{p.br.}$,
the electron and proton bremsstrahlung terms.

The total charge-even radiative correction factor is then:
\begin{eqnarray}\label{eq:deltaeven}
\sigma = \sigma_{Born} (1 + \delta_{even}) ~, \nonumber \\
\delta_{even} = 2 e^2 C_{even} / A_{1\gamma}^2 ~.
\end{eqnarray}
Two terms contribute to the charge asymmetry in elastic $e^\pm p$ scattering:
the interference between the Born and two-photon exchange diagrams and the
interference between electron and proton bremsstrahlung. Both of these terms
have infrared divergent contributions, but these divergences cancel in the sum
of the two contributions, making the QED description of the $e^\pm
p$-scattering self-consistent. This interference effect for the standard
kinematics of elastic $e^\pm p$-scattering experiments is dominated by
soft-photon emission and results in a factorizable correction already included
in the standard calculations of radiative corrections~\cite{mo69, tsai71,
ent01}.

The ratio of $e^\pm p$ scattering cross sections can thus be written as
follows:
\begin{eqnarray}
R = \frac{\sigma(e^+p)}{\sigma(e^-p)} \approx
\frac{1+\delta_{even}-\delta_{2\gamma}-\delta_{e.p.br.}}
     {1+\delta_{even}+\delta_{2\gamma}+\delta_{e.p.br.}} \nonumber \\
\approx  1 - 2 ( \delta_{2\gamma} + \delta_{e.p.br.})/(1+\delta_{even}) ~,
\label{eq:R}
\end{eqnarray} 
where $\delta_{even}$ is the total charge-even radiative correction
factor and $\delta_{2\gamma}$ and $\delta_{e.p.br.}$ are the
fractional TPE and lepton--proton interference contributions.  Note
that the sign of $\delta_{2\gamma}$ and $\delta_{e.p.br.}$ are chosen
by convention such that they appear as additive corrections for
electron scattering.  However, the sign of these corrections is
determined from the evaluation of the full expression given in
Eq.~\ref{eq:sqamptot}. Typically, a correction is applied to account
for the effect $\delta_{e.p.br}$ to isolate the TPE contribution:
\begin{equation}
\label{eq-R2g}
R_{2\gamma} \approx 1 - 2 \delta_{2\gamma} / (1+\delta_{even}).
\end{equation}
Where $R$ is the measured $e^+/e^-$ ratio and $R_{2\gamma}$ is the
ratio after applying corrections for the $e$--$p$ interference term.
The quantity $R_{2\gamma}$ corresponds to the quantity that is
typically quoted by such measurements, although the notation is not
always consistent.

Note that most previous extractions neglect the charge-even
contributions, assuming that $R = 1 - 2\delta_{2\gamma} -
2\delta_{e.p.br.}$ and $R_{2\gamma} = R + 2\delta_{e.p.br.} = 1 -
2\delta_{2\gamma}$.  Because the factor $\delta_{even}$ is typically
small (20--30\%) and negative, this means that assuming
$\delta_{2\gamma} = (1-R_{2\gamma})/2$ overestimates the TPE
contribution by approximately 20--30\%.  Because most extractions are
consistent with $\delta_{2\gamma}=0$, this rescaling of the TPE
contribution has minimal effect. More significant is the fact that
$\delta_{even}$ is neglected when applying the correction for the
$e$--$p$ interference term.  Because this correction is always a
reduction in the ratio, typically 1--5\%, this yields a systematic
underestimate of $R_{2\gamma}$ up to $\sim$1\%.

TPE corrections were extensively studied in the 1950s and 1960s. Early
calculations suggested that the contributions were
small~\cite{drell57, drell59, greenhut69}, and early measurements
comparing electron and positron scattering~\cite{yount62, browman65,
  anderson66, bartel67, cassiday67, anderson68, bouquet68, mar68,
  hartwig75} did not observe significant TPE contributions. Therefore,
most experiments have neglected the TPE corrections and typically
apply an uncertainty in the radiative correction procedure of roughly
1--1.5\%, dominated by the uncertainty in TPE corrections.

Two-photon exchange contributions have become a key issue in the field
in the last decade as a result of a significant discrepancy between
high-$Q^2$ polarization transfer measurements~\cite{punjabi05,
  puckett10, puckett11} of the proton form factor ratio $G_E^p/G_M^p$
and Rosenbluth separation extractions~\cite{andivahis94, christy04,
  qattan05} utilizing unpolarized scattering. Rosenbluth extractions
generally showed that both $G_E^p$ and $G_M^p$ approximately follow
the dipole form, $G_D = (1+Q^2/(0.71 {\rm~GeV}^2))^{-2}$, so that the
ratio $G_E^p/G_M^p$ is constant, while polarization measurements
showed the ratio decreasing linearly with $Q^2$.

One possible explanation of this discrepancy is a TPE contribution to the
cross section.  Explaining the difference between these techniques requires
an angle-dependent cross-section correction of 5--8\% at large
$Q^2$~\cite{arrington04d, borisyuk07b, belushkin08, qattan11a, qattan11b}.
However, this \textit{assumes} that the cross section change fully resolves
the discrepancy.  The form factor ratio discrepancy does not provide
significant cross section constraints at low $Q^2$.

Calculations of the TPE corrections were revisited~\cite{guichon03, blunden03,
chen04} in the wake of the form-factor discrepancy and initial calculations of
the TPE correction brought the Rosenbluth results into near agreement with the
polarization results. While low $Q^2$ calculations generally
agree~\cite{blunden05, borisyuk06, borisyuk07a, borisyuk08, borisyuk12,
arrington12d}, all of the available calculations have significant model
dependence at large $Q^2$.  While the hadronic calculations of Blunden,
Melnitchouk and Tjon~\cite{blunden03, blunden05} include only the proton
intermediate state, they fully reconcile the cross section and polarization
measurements up to $Q^2 \approx 2$~GeV$^2$ and resolve most of the discrepancy
at higher $Q^2$~\cite{arrington07c}. The effect of an intermediate $\Delta$
contribution in diagrams (e) and (f) of Fig.~\ref{fig:diagrams} has been
estimated and has a much smaller contribution, although it may have a more
significant contribution to the polarization measurements~\cite{kondratyuk05,
borisyuk12}. Calculations using a generalized parton distribution
formalism~\cite{chen04, afanasev05}, dispersion relations~\cite{borisyuk08},
and a QCD factorization approach~\cite{kivel09, borisyuk09} also yield TPE
contributions that can resolve a large part of the discrepancy.  Details of
these calculations and the issues involved can be found in recent
reviews~\cite{carlson07, arrington11b}.

Some limits for TPE contributions can be set based on existing cross
section and polarization measurements, combined with the known
properties of the OPE and TPE contributions.  In the Born
approximation, the reduced cross section depends linearly on $\epsilon
= (1+2(1+\tau)\tan^2(\theta/2))^{-1}$, where $\tau=Q^2/4M_p^2$.
Corrections beyond single-photon exchange can yield nonlinearities in
the reduced cross section, but existing data show that the corrections
are nearly linear~\cite{tvaskis06, tomasi05, chen07}. A recent
measurement of the $\epsilon$ dependence of the polarization
transfer~\cite{meziane10} also sets limits on TPE corrections, but the
precision is not sufficient to rule out the available calculations as
only the $\epsilon$ dependence, and not the overall size of the
extracted form factor, can be constrained.  In addition, even if there
is no contribution to the polarization transfer data, there can still
be a significant impact on the cross section~\cite{borisyuk11,
  guttmann11}.

It is clear that a direct confirmation of the presence of TPE
corrections is needed, as well as the data necessary to validate
calculations required for measurements that may be sensitive to TPE
effects. Since $\delta_{2\gamma}$ is expected to be on the order of a
few percent, one needs to measure $R$ to within an uncertainty of $\sim
1$\%.  While the early measurements found no significant TPE
contributions ~\cite{yount62, browman65, anderson66, bartel67,
  cassiday67, anderson68, bouquet68, mar68, hartwig75}, a combined
analysis of these earlier experiments, based on the idea that an
angle-dependent correction could reconcile the form factor
measurements, gave some evidence for such a
contribution~\cite{arrington04b}. The problem is that the low
luminosity of the secondary beams generally limited measurement to low
$Q^2$ values or small angle, where TPE contributions are not expected
to be larger than 1\%.

There are several recent attempts to improve on previous TPE
measurements comparing $e^\pm p$ and $\mu^\pm p$ scattering.  Two of
these are straightforward experimentally, utilizing electron and
positron beams from the VEPP-3 storage ring~\cite{VEPP-3, nikolenko10,
  gramolin12} or the DORIS ring at DESY~\cite{olympus, kohl11}.  The
storage rings allow for good control of systematics, though the
available luminosities limit the measurements to be done at lower beam
energies and thus lower $Q^2$ values and also limit the statistical
precision of the data at small $\epsilon$ where TPE contributions are
believed to be large. The
MUSE Collaboration~\cite{museproposal} has proposed to compare $e^\pm
p$ and $\mu^\pm p$ scattering at very low $Q^2$. This is motivated by
the ``proton radius puzzle''; the difference between proton radius
extractions involving muonic hydrogen~\cite{pohl10} and those
involving electron--proton interactions~\cite{mohr08, bernauer10,
  zhan11}. The MUSE experiment will compare electron and muon
scattering to look for indications of lepton non-universality, but
will also examine TPE corrections, which are important in the radius
extraction from electron scattering data~\cite{rosenfelder00,
  blunden05, arrington11c, bernauer11}.  

We have taken a very different approach to comparing $e^+ p$ and $e^-p$
scattering.  Rather than alternating between mono-energetic $e^+$ and $e^-$
beams, we generate a combined beam of positrons and electrons covering a
range of energies and use the large-acceptance CLAS spectrometer in Hall B
of Jefferson Lab to detect both the scattered lepton and struck proton.
The kinematics for elastic scattering are overconstrained in such a
measurement, allowing us to reconstruct the initial lepton energy, as well
as ensuring that the scattering was elastic.  This allows a simultaneous
measurement of electron and positron scattering, while also covering a
wide range in $\epsilon$ and $Q^2$.  As such, the full experiment utilizing
the setup described here~\cite{prop06} is the only TPE measurement that will
extract the $\epsilon$ dependence of the TPE corrections at fixed $Q^2$, such
that they can be directly applied to Rosenbluth separations of the form
factors.

\section{Experimental Details}
\label{Sec:ExptDetails}

This section and the next describe the novel technique used to create a mixed
electron-positron beam, the methods for extracting elastic scattering
events using this beam, and the initial measurement of the
positron-electron elastic scattering ratio over a narrow kinematic
range.

\begin{figure}[thb]  
\begin{center} 
\includegraphics[width=0.49\textwidth]{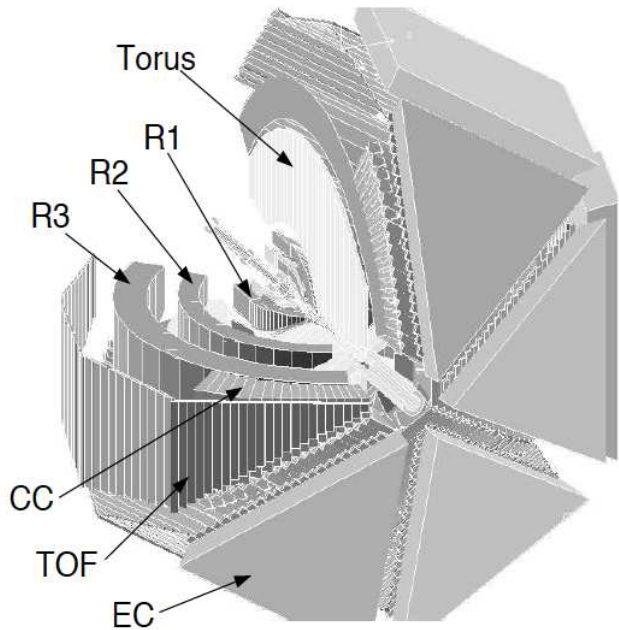}
\caption{\label{fig:clas} 
Three dimensional view of CLAS showing the beamline, drift chambers (R1, R2,
and R3), the Cherenkov Counter (CC), the Time of Flight system (TOF) and the
Electromagnetic Calorimeter (EC).  In this view, the beam enters the picture
from the upper left corner.}
\end{center} 
\end{figure}

The experiment took place at the Thomas Jefferson National Accelerator
Facility (Jefferson Lab) and used the CEBAF Large Acceptance
Spectrometer (CLAS) \cite{clasNIM} in Hall B to detect scattered
particles.  CLAS (see Fig.~\ref{fig:clas}) is a nearly 4$\pi$
detector. The magnetic field is provided by six superconducting coils
that produce an approximately toroidal field in the azimuthal
direction around the beam axis.  The regions between the six magnet
cryostats are instrumented with identical detector packages called
sectors. Each sector consists of three regions of drift chambers (R1,
R2, and R3) to determine the trajectories of charged particles
\cite{clasDCNIM}, Cherenkov Counters (CC) for electron identification
\cite{clasCCNIM}, scintillation counters for Time of Flight (TOF)
information \cite{clasSCNIM}, and Electromagnetic Calorimeters (EC)
for electron identification and neutral particle detection
\cite{clasECNIM}. The R2 drift chambers are in the region of the
magnetic field and provide tracking that is then used to determine
particle momenta with $\delta p/p\sim 0.6$\% This experiment did not
use the CC and used the EC only in the trigger.

\begin{figure*}[tbp]
\begin{center}
\includegraphics[width=0.95\textwidth]{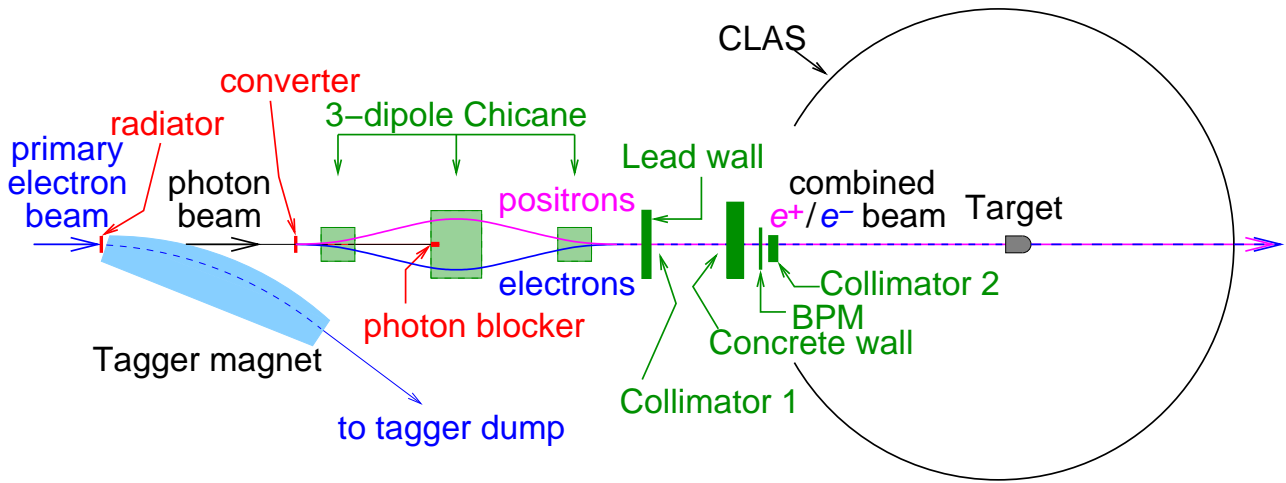}
\caption{(Color online) Beamline sketch for the CLAS TPE
  experiment. Shielding elements around the chicane and tagger are not
  shown. The chicane bends the electron and positron trajectories in
  the horizontal plane, not the vertical plane.  The electron and
  positron directions are selected by the chicane polarity.  Drawing
  is not to scale.}
\label{fig:TPEsketch}
\end{center}
\end{figure*}

We produced a simultaneous mixed beam of electrons and positrons by
using the primary electron beam to produce photons and then using the
photon beam to produce $e^+e^-$ pairs (see
Fig.~\ref{fig:TPEsketch}). A 40--80 nA 3.3 GeV electron beam struck a
$4.5\times10^{-3}$ radiation-length (RL) gold radiator to produce a
bremsstrahlung photon beam.  The electrons were diverted by the Hall B
tagger magnet \cite{clastagger} into the standard underground beam
dump.  The photon flux was about $10^2$ greater than previous Hall B
photon fluxes, requiring substantial additional shielding around the
beam dump.

The photon beam passed through a 12.7 mm diameter nickel collimator and
then struck a $5.1\times10^{-2}$ RL gold converter to produce
$e^+/e^-$ pairs.  The combined photon-lepton beam then entered a
three-dipole chicane to horizontally separate the electron, positron and photon
beams.  The photon beam was stopped by a tungsten block in the middle
of the second dipole.  The lepton beams were recombined into a single
beam by the third dipole, which then proceeded to a liquid hydrogen
target at the center of CLAS.  Fig.~\ref{fig:TPEsketch} shows the
layout of the beamline and Table~\ref{tab:beam} lists the relevant
parameters.

\begin{table}[htb]
 \begin{center}
 \begin{tabular}{|c|c|}\hline
Primary Beam & $40\leq I\leq 80$ nA \\
             & $E=3.3$ GeV   \\ \hline
Radiator (gold) & $4.5\times 10^{-3}$ RL \\
Photon Collimator  & 12.7  mm ID \\
                              & length = 30 cm \\ \hline
 Converter (gold) & $5.1\times 10^{-2}$ RL \\ \hline
Italian Dipole & $B\approx 0.4$ T \\
 & $L\approx 0.5$ m \\ \hline
PS Dipole & $B\approx 0.38$ T \\
 & $L\approx 1$ m \\ \hline
 Lepton Collimator 1 (lead) & 2.5 cm ID \\ \hline
Fiber BPM   & 8 cm $\times$ 8 cm \\ \hline
Lepton Collimator 2 (lead) & 6 cm ID \\ \hline
{LH2 target}    & diameter=6 cm \\
                      &length=18 cm \\ \hline
CLAS Torus Current & $\pm 1500$ A \\ \hline
\end{tabular}
 \end{center}
 \caption{Running conditions. ID = Inner Diameter, RL=radiation lengths.}
\label{tab:beam}
\end{table}

The TPE chicane consisted of three dipole magnets.  The first and
third dipoles were the so-called ``Italian Dipoles'' and the second
dipole was the pair spectrometer magnet (PS).  The Italian Dipoles
were operated with a magnetic field of $B \approx \pm0.4$ T and were
about 0.5~m long.  They were powered in series by a single power
supply.  The PS had a field of $B\approx \mp 0.38$ T and was about 1~m
long. The oppositely charged leptons were separated horizontally and
recombined by the chicane. The photon beam was absorbed by a 4-cm wide
and 35-cm long tungsten photon blocker positioned with its upstream
face at the entrance aperture of the PS magnet.

The momentum acceptance of the chicane is determined by the width of
the photon blocker and the apertures of the PS.  The width of the
photon blocker ($\pm$0.02~m) determined the maximum lepton momentum
and the PS aperture of approximately $\pm0.2$ m determined the minimum
lepton momentum. Because the three dipoles are left-right symmetric,
the two lepton beams should be identical.  The final lepton beam
energy ranged from approximately 0.5 to 3.2 GeV.

Either of the separated lepton beams could be blocked by inserting one
of the two ``beam blockers'' as the beam was diverging. These are
standard-sized lead bricks that could be inserted at the exit of the
first Italian Dipole. They were used to block either lepton beam to
allow study of the other beam by itself.

We built two large shielding structures (not shown in
Fig.~\ref{fig:TPEsketch}), one between the first and second chicane
magnets and one between the second and third chicane magnets.  This
shielding reduced the background produced by the chicane that would
otherwise strike the CLAS detectors.  We placed a 1-m by 1-m by 0.1-m
thick lead wall immediately after the chicane to further reduce
chicane backgrounds.  This wall had a 2.5 cm diameter collimator
(``collimator 1'') to allow the mixed lepton beams through.  We added
other shielding, including a large concrete wall approximately 2 m
upstream of the entrance to CLAS.

We measured the position of the lepton beams using a
scintillating-fiber beam position monitor (BPM).  This (BPM) is an
array of $16\times 16$ scintillating fibers read out by a multichannel
PMT and was located approximately 15 cm upstream of the entrance to
CLAS.  The fibers were 1 mm $\times$ 1 mm and had a spacing of 5
mm. We used the BPM to monitor the width and position of the lepton
beams continuously during the measurement.  A second collimator with a
6-cm diameter aperture was positioned right after the BPM.

This experiment was primarily an engineering test run to determine the
feasibility of using this mixed electron-positron beam line to
definitively measure the ratio $R=\sigma(e^+p)/\sigma(e^-p)$
to resolve the proton form factor discrepancy. Prior to the data
taking phase of the test run we varied the experimental parameters to
optimize the lepton beam luminosity.  The beam luminosity was limited
by requiring that the CLAS drift chamber occupancy in each
sector and each region all be less than 3\%.  We varied the incident
beam current, the radiator thickness, the photon collimator diameter,
the converter thickness, the chicane magnet currents, and the first
lepton collimator diameter.  We also greatly improved the shielding.
The values listed in Table \ref{tab:beam} show the final optimized
values.

\begin{figure}[b]
\begin{center}
\includegraphics[width=0.49\textwidth]{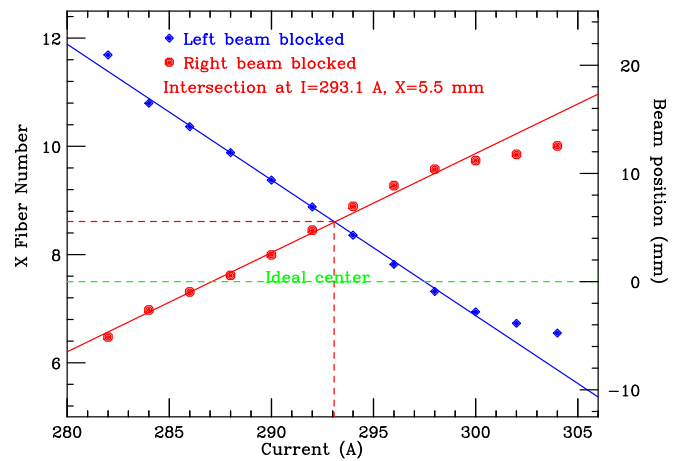}
\caption{(Color online) Position of the individual lepton beams as a
  function of the current in the first and third dipoles. Data points
  are measured beam centroid positions at the fiber detector and the
  lines are fits to points 2--10.}
\label{fig:xscan}
\end{center}
\end{figure}

For example, in order to make sure that both lepton beams had the same
centroid position, we varied the current in the Italian Dipole magnets
while keeping the PS dipole current fixed.  We blocked one lepton beam
and measured the position of the centroid of the other beam as a
function of the Italian Dipole current.  We then blocked the other beam
and repeated the measurement for the first beam. Fig.~\ref{fig:xscan}
shows the results. We fit straight lines to the linear parts of the
beam position vs.~magnet current data.  The intersection of the two
linear fits indicates the Italian Dipole magnet current that optimized
the centering of the two lepton beam spots. This optimized centering
is approximately 5~mm off the expected beam center indicating a likely
misalignment of the BPM.

\begin{figure}[hbtp]
\rotatebox{270}{\includegraphics[width=0.3\textwidth,clip=true]{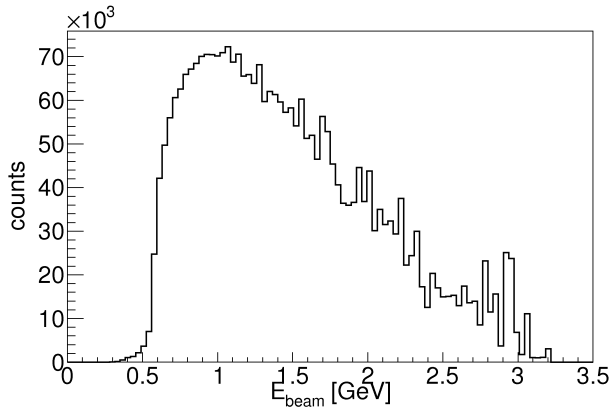}}
\begin{center}
  \caption{Reconstructed beam energy (electrons and positrons
    combined) at the target for elastic events detected in CLAS.  Each
    event is weighted by one over the elastic cross section to recover
    the initial beam energy distribution.}
\label{fig:BeamSim}
\end{center}
\end{figure}

The reconstructed beam energy distribution for elastic scattering
events detected in CLAS is shown in Fig.~\ref{fig:BeamSim}.  The shape
of this distribution is a convolution of the incident beam energy
distribution, the elastic scattering cross section and the CLAS
acceptance. The maximum flux is at low energies, consistent with the
bremsstrahlung cross section. This feature limits the measurement here
to low $Q^2$ and high $\epsilon.$ We estimate the integrated beam current
of each beam to be on the order of 1 pA. The width of the beam varied
as a function of energy from an RMS of 1.6 cm for beam energies in the
range $0.5\leq E_{beam} \leq 1.0$ GeV to 1.1~cm for $2.5 \leq E_{beam}
\leq 3.0$ GeV. The lepton beam was incident on a 6-cm diameter
cylindrical target of liquid hydrogen.  The target had hemispherical
endcaps, kapton walls, and a total length of 18 cm.

The scattered leptons and protons were detected in the CLAS
spectrometer.  The CLAS event trigger required a particle to deposit
some energy in the EC in any sector (the threshold was chosen to
accept minimum ionizing particles) and a hit in a TOF counter in the
opposite sector.

The magnetic fields of the CLAS torus magnet and the beamline chicane
can be reversed periodically to reduce lepton charge-dependent
experimental asymmetries.  However, during this run only the CLAS
torus field was reversed. By using simultaneous mixed
electron-positron beams we eliminated the effect of time-dependent
detector efficiencies. By taking data with both chicane polarities, we
would eliminate, within uncertainties, any flux-dependent differences
between the left and right beams.

\begin{figure}[htbp]
\begin{center}
\rotatebox{270}{\includegraphics[width=0.32\textwidth,clip=true]{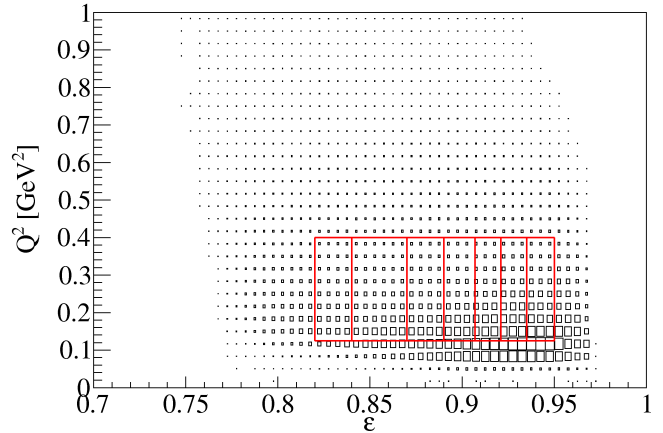}}
\caption[Binning scheme.]{(Color online) The experimental acceptance
  in momentum transfer and virtual photon polarization shown for
  negative torus polarity and both lepton charges. The red boxes show
  the binning for the data presented here.}
\label{bins}
\end{center}
\end{figure}

The data were taken as part of a test run to verify the feasibility of
the experiment.  The test run was sufficiently successful that we took
about 1.5 days of experimental data after commissioning the $e^+e^-$
beamline. These data allowed us to test our data analysis techniques
and to measure the ratio $R$ at low $Q^2$ and high $\epsilon$ as shown
in Fig.~\ref{bins}.

\section{Data Analysis}
\label{sec-analysis}

This analysis confronted us with a number of issues uncommon to other
CLAS experiments.  The primary problems were (a) determining the
energy of the incident lepton, (b) making the analysis
lepton-charge-independent, and (c) identifying the lepton and proton
without using the Cherenkov Counter because its efficiency depends on
whether the lepton is in-bending or out-bending.  Our solution to this
was to require the detection of both the proton and the lepton in each
event, to exploit the restricted kinematics of elastic-scattering to
identify elastic events, and to match the detector acceptances for the
two types of events (electron-proton and positron-proton).  A
description of the important analysis procedures is given below.

\subsection{Elastic Event Identification}
\label{sec-eventID}

We first selected events with only two detected charged particles in
opposite sectors and where their charges were either positive/negative
or positive/positive.

For positive/positive events, we used information from the TOF
counters to determine which particle was the proton and which was the
positron.  We initially identified the positron by requiring that
$\beta = v/c >0.9$, noting that at these kinematics protons have
$\beta<0.9$.  With this loose PID cut in place we then verified the
PID assignment by following the rest of the elastic event
identification chain. If the event did not subsequently satisfy the
elastic scattering cuts listed below, we swapped the identities of the
two positive particles and checked to see if the event then satisfied
the elastic cuts. None of these swapped ++ events passed the elastic
cuts, indicating that the $\beta>0.9$ requirement properly identified
positrons in all cases.

The additional cuts included fiducial cuts, bad detector removal,
event vertex cuts, and four elastic scattering kinematic cuts.  These
are summarized in the list below. As will be shown, these cuts were
correlated in that any single cut had minimal effect when all of the
other cuts are applied.  This leads to very clean elastic event
distributions with minimal background contamination.
Figs.~\ref{fig:phicut}--\ref{fig:theta} show some of the variables
before and after the elastic cuts.  They are shown for all four
combinations of torus magnet polarity and lepton charge.  Unless
otherwise indicated, they show the combined data for both torus
polarities.
\begin{enumerate}
\item {\bf Bad paddle removal.}  As CLAS has aged, some of the TOF
  detector photomultiplier tubes (PMT) have deteriorated and have low
  gain leading to very poor efficiency.  Events with particles that
  hit one or more of these detectors were removed from the analysis.
\item {\bf $Z$-vertex.} The particle origin along the beamline
  ($z$-vertex) was reconstructed as part of the trajectory
  measurement.  A cut was placed on $z$-vertex to ensure that events
  came from the LH$_2$ target.
\item {\bf Distance of closest approach between lepton and proton
    candidates.} This is defined as the distance between the two
  tracks (lepton and proton) at their closest point.  A cut was placed
  on this distance to ensure the two tracks came from the same
  interaction.
\item {\bf Fiducial cuts.} Fiducial cuts in angle and momentum were
  used to select the region of CLAS with uniform acceptance for both
  lepton polarities, thus matching the acceptances for electron and
  positron.
\item {\bf Azimuthal opening angle (co-planarity).}  Since there are
  only two particles in the final state, these events must be
  co-planar.  Fig.~\ref{fig:phicut} shows the azimuthal-angle
  difference between events before and after all other cuts.
\begin{figure}[htb]
  \begin{center}
    \includegraphics[width=0.48\textwidth,height=7cm]{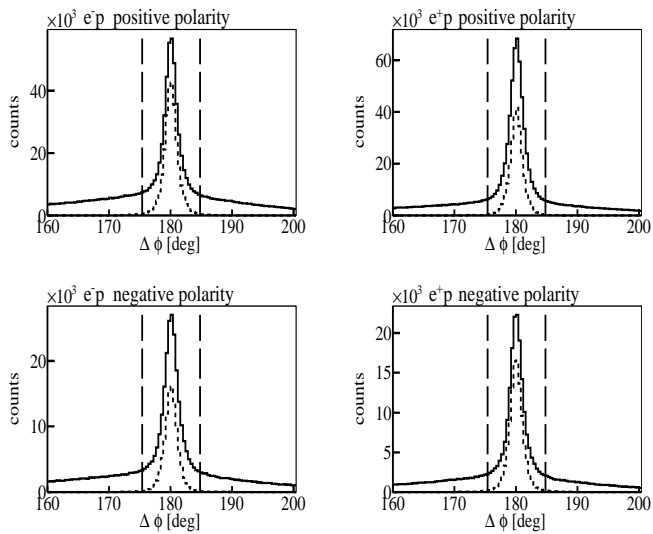}
    \caption{Angle between lepton and proton ($\Delta\phi$)
      distributions for event type and torus polarity as indicated.
      The solid histogram is the data with only the opposite sector
      cut. The dotted histogram is after all other cuts. The dashed
      lines show the $\Delta\phi$ cut.}
    \label{fig:phicut}
  \end{center}
\end{figure}
\item {\bf Transverse momentum.} Conservation of momentum requires the
  total transverse momentum, $p_t$ (with respect to the incident beam)
  of the final-state elastic scattering products to be zero. See
  Fig.~\ref{fig:transp}.
\begin{figure}[htb]
  \begin{center}
    \includegraphics[width=0.48\textwidth,height=7cm]{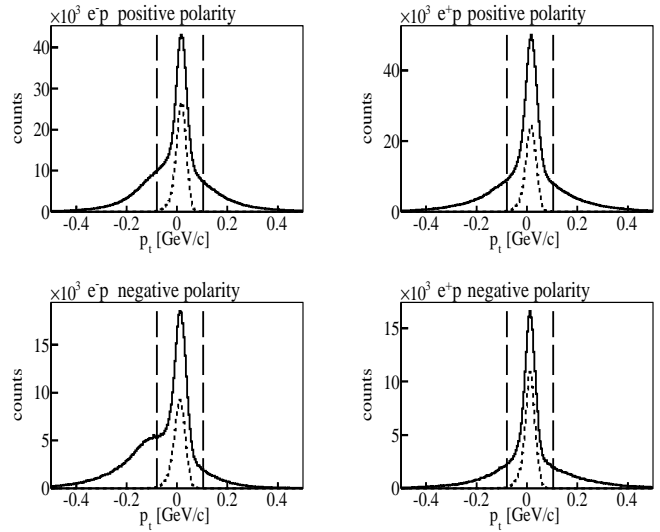}
    \caption{Reconstructed transverse momentum distributions for event
      type and torus polarity as indicated.  The solid histogram is
      the data with only the opposite sector cut.  The dotted
      histogram is after all other cuts. The dashed lines show the
      $p_t$ cut.}
    \label{fig:transp}
  \end{center}
\end{figure}
\item {\bf Beam energy difference.} Because we measured the 3-momenta
  for both particles in the final state, our kinematics are over
  constrained. This allows us to reconstruct the unknown energy of the
  incident lepton in two different ways. Eq.~\ref{ebeam1} calculates
  the incident energy using the scattered lepton and proton angles,
  whereas Eq.~\ref{ebeam2} calculates this from the total momentum
  along the $z$-direction.
\begin{eqnarray} 
  E_{beam}^{angles} &=& m_p \left(\cot\frac{\theta_{e}}{2}\cot \theta_p - 1
  \right) 
  \label{ebeam1}\\ 
  E_{beam}^{mom}&=& p_{e}\cos \theta_{e} + p_p\cos \theta_p 
  \label{ebeam2}
\end{eqnarray} 
For elastic scattering events, these two quantities should be equal.
We cut on $\Delta E_{beam} = E_{beam}^{angles}-E_{beam}^{mom}$ (see
Fig.~\ref{fig:dEbeam}).  The 7 to 22 MeV shift in the centroids from
zero is due to particle energy loss in the target, which reduces the
value of $E_{beam}^{mom}$.  Note that we used $E_{beam}^{angles}$ in
all calculations that depend on beam energy (e.g., $Q^2$, $W$,
$\epsilon$).
\begin{figure}[htb]
  \begin{center}
    \includegraphics[width=0.48\textwidth,height=7cm]{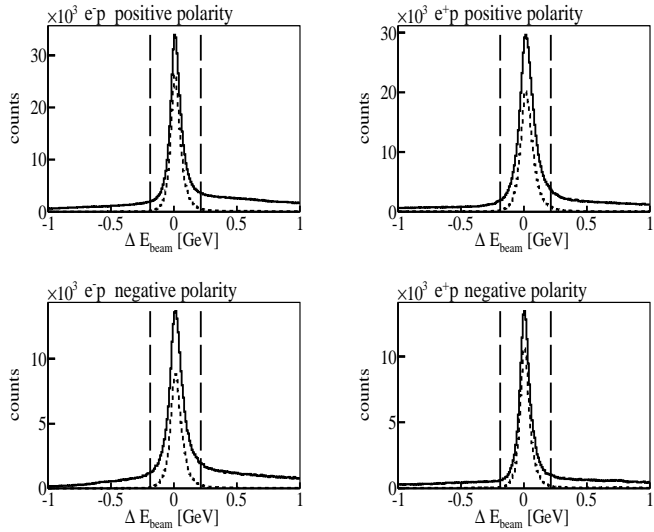}
    \caption{$\Delta E_{beam}$ for event type and torus polarity as
      indicated.  The solid histogram is the data with only the
      opposite sector cut. The dotted histogram is after all other
      cuts.  The dashed lines show the $\Delta E_{beam}$ cut.}
    \label{fig:dEbeam}
  \end{center}
\end{figure}
\item {\bf Momentum polar angle.} We cut on the polar angle of the
  final-state total-momentum ($\vec P_f = \vec p_{l'} + \vec p_{p'}$)
  with respect to the $z$-axis.  Deviations from zero may be due to
  inelastic events, mis-reconstructed scattered particles, or
  multiple-scattered final particles.  To discard these background
  events, we required $\theta_{P_f} < 5^\circ$ (see
  Fig.~\ref{fig:theta}).  This cut is largely redundant to the
  transverse momentum cut.
\begin{figure}[hbtp]
  \begin{center}
    \includegraphics[width=0.48\textwidth,height=7cm]{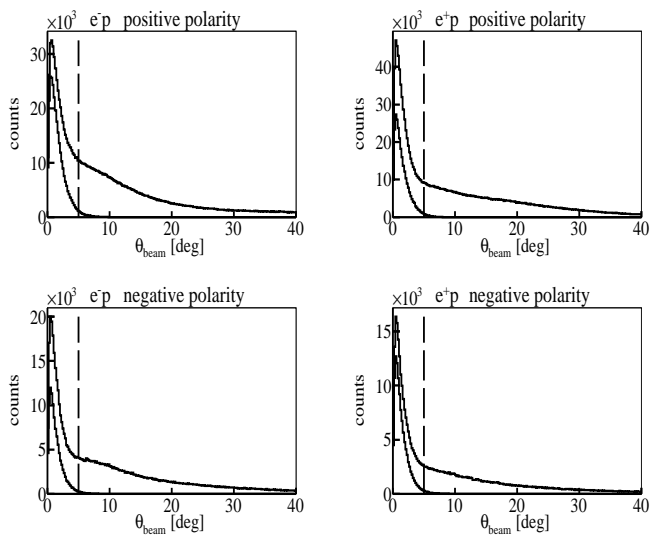}
    \caption{Reconstructed polar angle of the beam for event type and
      torus polarity as indicated.  The solid histogram is the data with
      only the opposite sector cut and the dotted histogram is after
      all other cuts.  The dashed lines show the $\theta_{P_f}$ cut.}
    \label{fig:theta}
  \end{center}
\end{figure}
\end{enumerate}

The cuts for items 3, 5, 6, 7 and 8 were determined by fitting a
Gaussian to the peak of the combined distribution for that variable, including
both event types ($e^+p$ and $e^-p$) and both torus polarities and setting the cut to $\pm
4\sigma$.  The widths of these distributions did not depend
significantly on either torus polarity or event type.

\begin{figure}[htb]
  \begin{center}
    \includegraphics[width=0.48\textwidth,height=7cm]{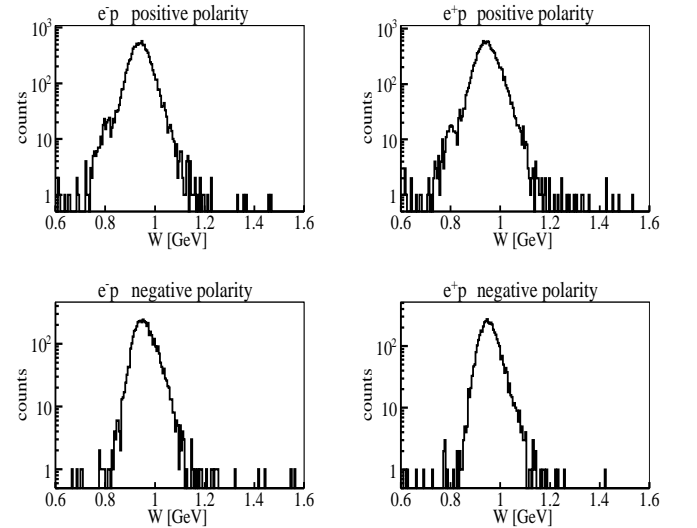}
    \caption{Top two panels show the $W$ distributions (in GeV) for
      positive torus polarity electron (left) and positron (right)
      events for $0.820\le\epsilon\le 0.840$ and $\langle Q^2\rangle
      =0.206$ GeV$^2$ with all cuts applied.  The bottom two panels
      show the same for negative torus polarity events.}
    \label{fig:W1}
  \end{center}
\end{figure}

The cleanliness of the final data sample after these cuts were applied is
shown in Fig.~\ref{fig:W1}, which shows the invariant mass of the
virtual photon plus target proton, $W=\sqrt{m_p+2m_p\nu-Q^2}$,
distribution for one of our bins in $\epsilon$.  The peak is at the proton
mass and shows virtually no hint of non-elastic background. Using side
bands on either side of the peak we estimate the background to be in
the range of 0.3 to 0.4\%.  Since the background is equal to within
uncertainties for both lepton species the effect on $R$ is negligible.

The distribution of elastic events in $Q^2$ vs.~$\epsilon$ after all
cuts is shown in Fig.~\ref{bins}.  The boxes in the figure show the
bins used for this analysis. The final results cover a single $Q^2$
bin ($0.125\leq Q^2\leq 0.400$ GeV$^2$ with $\langle Q^2\rangle
=0.206$ GeV$^2$) and seven bins in $\epsilon$ ($0.830\leq \epsilon\leq
0.943$) such that we have similar statistical uncertainties in each
$\epsilon$ bin.

\subsection{Acceptance Matching and Corrections}
\label{sec-acceptance}

In order to calculate the ratio $R=\sigma(e^+p)/\sigma(e^-p)$, we must
ensure that the detector acceptance does not depend on lepton
charge.  We first calculate $R$ by calculating the ratio of $e^+p$
to $e^-p$ events for a given torus polarity.  In this ratio, the
proton acceptance cancels. However, the CLAS acceptances for electrons
and positrons for a given kinematic bin differ because one bends away
from the beamline while the other bends toward the beamline in the
CLAS magnetic field.  

We have accounted for acceptance differences in three steps.
First, we match acceptances by using the fiducial cuts to select
regions in ($p,\theta,\phi$) space where CLAS is almost 100\%
efficient in detecting both electrons and positrons.  Second, we
correct for differences due to dead detectors using a ``swimming''
algorithm to check whether an $e^+p$ would have been detected had it
been an $e^-p$ event (and vice versa).

For example, if an $e^+p$ event is detected and passes all the elastic
cuts, then the swimming algorithm generates a conjugate lepton, in
this case an $e^-$, with the same momentum $\vec p$ as the $e^+$ and
calculates (``swims'') its trajectory through the CLAS detector system
and magnetic field.  If the conjugate lepton falls within the CLAS
acceptance, then the original event is kept. If the conjugate lepton
falls outside of the CLAS acceptance (either outside fiducial cuts or
hits a dead paddle), the event is discarded.

In the third step, any remaining acceptance
differences can be removed by measuring $R$ for both torus polarities
and constructing a double ratio.  The number of detected elastic
events for a given torus polarity ($t=\pm$) and a given lepton charge
($l=\pm$) should be proportional to the cross section times the
unknown torus-polarity-related and lepton-charge-related detector
efficiency and acceptance function $f_t^l$:
\[
N_t^l\propto \sigma(e^lp)f_t^l.
\]
Thus, for one torus polarity, the simultaneously measured ratio $R_t$ will be
\[
R_t = \frac{N_t^+}{N_t^-} = \frac{\sigma(e^+p)f_t^+}{\sigma(e^-p)f_t^-} \quad .
\]
Taking the square-root of the product of the single-polarity ratios we
get
\begin{eqnarray}
  R&=&\sqrt{R_+R_-} = \sqrt{\frac{N_+^+ }{N_+^- }\cdot \frac{N_-^+}{ N_-^-} } \nonumber \\
  &=& \sqrt{\frac{\sigma(e^+p)f_+^+}{\sigma(e^-p) f_+^-} \cdot
    \frac{\sigma(e^+p)f_-^+}{\sigma(e^-p) f_-^-} } \nonumber \\
  &=& \frac{\sigma(e^+p)}{\sigma(e^-p)} \label{eq:Rfinal},
\end{eqnarray}
where by charge symmetry, one expects $f^+_+=f^-_-$ and $f^+_-=f^-_+$.
The unknown lepton acceptance functions are expected to cancel in the
double ratio. The proton acceptance cancels out independently in the
single ratios $R_+$ and $R_-$.

We checked the quality of the corrections described above by comparing
it to two other methods.  The first approach is to apply an acceptance
correction to the $e^+$ and $e^-$ data separately based on a full
Monte Carlo (MC) study using GSIM, the GEANT-based CLAS Monte Carlo
simulation that included all dead detectors.  The second approach was
to calculate the double ratio with no acceptance corrections at all
since, in principle, all acceptances cancel out in the double-ratio.
We found the differences among the three values of $R$ to be smaller
than their statistical uncertainties.  We used the difference between
our swimming results and our MC-corrected results to estimate the
dead-detector-related systematic uncertainty.

\subsection{Systematic Uncertainties}
The four major categories of systematic uncertainties in this analysis are:
\begin{enumerate}
\item{\bf Luminosity differences between electrons and positrons.} In
  this test run we could not independently measure the lepton beam
  luminosities and we did not have the time to take data for both
  polarities of the beam-line chicane magnetic field. Therefore we
  determined the relative luminosity uncertainty by a detailed
  GEANT4-based Monte Carlo study of the beam line that included all
  known lepton interactions.  The MC study showed that the relative
  flux difference between positrons and electrons on the target was
  less than 1\% for an ideal beamline.  Based on survey results, the
  alignment of beam line elements was within 1 mm.  The MC beam-line
  simulation showed that, for a single chicane polarity, a 1-mm
  change in the relative position of the collimators and magnets
  leads to a 5\% change in the electron-positron luminosity ratio.
\item{\bf Effects of elastic event ID cuts.} This was studied by
  varying the widths of these cuts (from the nominal 4-$\sigma$ cut to
  a 3-$\sigma$ cut or removing the cut entirely).  The differences in
  the final double-ratio results between the nominal and the varied
  cuts result in an estimated absolute uncertainty in $R$ of 0.0040.
\item{\bf Effects of fiducial cuts.} We also varied the cuts that
  define the good region of CLAS, again comparing the nominal
  double-ratio results to those with the varied fiducial cuts.  The
  estimated absolute uncertainty in $R$ is 0.0011.
\item{\bf Acceptance (dead detector) corrections.} As previously
  mentioned, this was done by comparing our nominal double-ratio
  results (using ``swimming'') to results using a MC correction.  The
  estimated absolute uncertainty in $R$ is 0.0071 and is the largest
  of our point-to-point systematic uncertainties.
\end{enumerate}

The 5\% luminosity-related uncertainty is a scale type uncertainty affecting
all points in the same way.  The other three items represent uncorrelated
point-to-point uncertainties, which added in quadrature give an
overall uncertainty of 0.0083 in $R$.

\section{Results}
\label{sec-results}

\begin{figure}[hb]
\begin{center}
\includegraphics[width=0.48\textwidth]{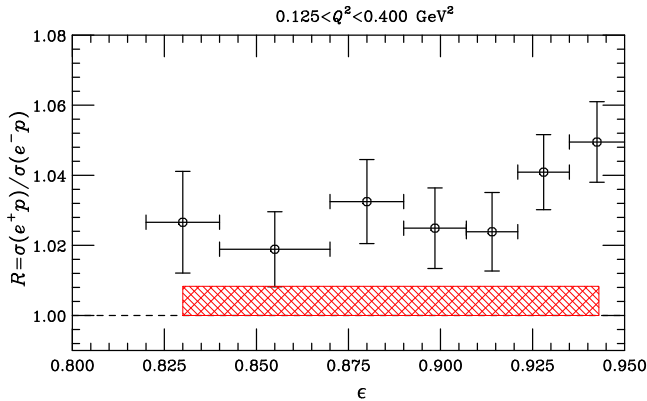}
\caption{(Color online) Measured ratio $R$ for acceptance matched data at
  $\langle Q^2\rangle=0.206$ GeV$^2$ before radiative
  corrections. Vertical error bars are statistical only and horizontal
  error bars show the range of the bin.  The red shaded band indicates
  the point-to-point systematic uncertainty ($1\sigma$) of the present
  data. The 5\% luminosity-related systematic uncertainty is not
  shown.}
\label{fig:Swim}
\end{center}
\end{figure}

\begin{figure}[t]
\begin{center}
\includegraphics[width=0.49\textwidth]{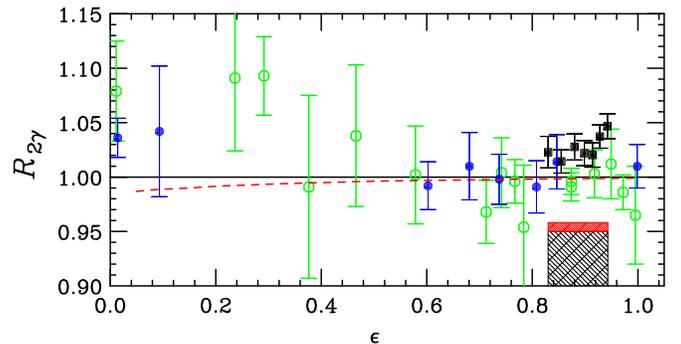}
\caption{(Color online) Ratio $R_{2\gamma}$ overlaid on the world
  data. Black filled squares are from this experiment at $\langle
  Q^2\rangle=0.206$ GeV$^2$ and have had radiative corrections
  applied, blue filled circles are previous world data at similar
  $Q^2$, and green hollow points the rest of the previous world data
  with $Q^2<2$ GeV$^2$ \cite{arrington04b}. The red densely shaded
  band indicates the point-to-point systematic uncertainty ($1\sigma$)
  and the black shaded band represents the scale-type systematic
  uncertainty (due to relative luminosity) on the present data.  The
  red dashed curve is the BMT calculation \cite{blunden05} at
  $Q^2=0.2$ GeV$^2$.}
\label{fig:Rcompare}
\end{center}
\end{figure}

Our final results are presented in Table~\ref{tab:Results} and
Fig.~\ref{fig:Swim}.  Fig.~\ref{fig:Rcompare} and
Table~\ref{tab:Results} show the results for $R_{2\gamma}$
(Eq.~\ref{eq-R2g}) after correcting the measured ratio $R$ for the
lepton-proton bremsstrahlung interference \cite{mo69}.  The
corrections reduce the measured ratio by 0.0049 at $\epsilon=0.830$ 
and decrease gradually to 0.0034 at $\epsilon=0.943$. The uncertainty 
on this correction is 0.0008 and is a combination of the uncertainty
in the cut-off value of $W$ used in the correction calculation and the
uncertainty in the term $\delta_{even}$.  This uncertainty is far
smaller than our other systematic uncertainties and can be
ignored. The average of our results, with the point-to-point
systematic uncertainty combined in quadrature with the statistical
uncertainty, is 1.027$\pm 0.005\pm 0.05$, with the last uncertainty
being due to the luminosity uncertainty. 

These measurements cover a very narrow range in $\epsilon$.
$R_{2\gamma}$ is not expected to vary over this narrow range of
$\epsilon$, especially at this low momentum transfer.  For example,
see the BMT calculation \cite{blunden05} shown in
Fig.~\ref{fig:Rcompare}. Therefore, the variation of these data
should be consistent with its uncertainties.  The standard deviation
of the seven data points is 0.01, which is consistent with and
slightly smaller than the statistical plus point-to-point
uncertainties.

We compare our results with the world's data at a similar value of
$Q^2$ as a function of $\epsilon$ in Fig.~\ref{fig:Rcompare}.  There
are seven previous data points in this range of $Q^2$.  Our data are
compatible with these points, although with significantly smaller
statistical uncertainties.  However, the 5\% systematic uncertainty
due to the luminosity prevents us from extracting any significant
conclusions about the size of the TPE effect.

\begin{table}[hbtp]
\begin{center}
\begin{tabular}{|c|c|c|c|c|c|c|}\hline
$\langle Q^2\rangle$ (GeV$^2$) & $\langle\epsilon\rangle$ & $R$  &
$R_{2\gamma}$ & $\delta R_{stat}$ & $\delta R_{sys}$ & $\delta R_{lum}$\\
\hline
      & 0.830 & 1.027 & 1.023 & 0.015 & &\\
      & 0.855 & 1.019 & 1.014 & 0.011 & &\\
      & 0.880 & 1.033 & 1.028 & 0.012 & &\\
0.206 & 0.899 & 1.025 & 1.022 & 0.012 & 0.0083 & 0.05\\
      & 0.914 & 1.024 & 1.020 & 0.011 & &\\
      & 0.928 & 1.041 & 1.037 & 0.011 & &\\
      & 0.943 & 1.050 & 1.047 & 0.012 & &\\ \hline
\end{tabular}
\caption{Charge asymmetry ratio and uncertainties.  $\langle
  Q^2\rangle$ and $\langle\epsilon\rangle$ show the average momentum
  transfer and photon polarization for that bin respectively,
  $R$ and $R_{2\gamma}$ show the measured value of
  $R=\sigma(e^+p)/\sigma(e^-p)$ before and after radiative corrections
  respectively, $\delta R_{stat}$, $\delta R_{sys}$ and $\delta
  R_{lum}$ show the statistical uncertainty, the point-to-point
  systematic uncertainty and the luminosity-related systematic
  uncertainty respectively.}
\label{tab:Results}
\end{center}
\end{table}

\section{Conclusion and Future Prospects}
\label{Raue:Conclusion}
We have presented a new technique for producing a mixed
electron-positron beam using bremsstrahlung to produce a secondary
photon beam from the primary electron beam and then pair-production to
produce a tertiary electron-positron beam from the photon beam.  We
aimed this beam at a liquid hydrogen target in the center of the CLAS
spectrometer.  We have presented analysis techniques to cleanly
extract elastic-scattering electron-proton and positron-proton events
and to minimize the charge-dependent experimental asymmetries.

We then used these techniques to extract
$R=\sigma(e^+p)/\sigma(e^-p)$, the ratio of positron-proton to
electron-proton elastic scattering cross sections over a limited
kinematic range at large $\epsilon$ and small $Q^2$.  The extracted
ratio is consistent with the world's data.  This ratio $R$ is directly
related to the magnitude of the Two Photon Exchange contribution to
electron-proton elastic scattering.

During late 2010 and early 2011 we conducted the full CLAS TPE
experiment using an incident beam energy of 5.5 GeV and significantly
greater luminosity. This experiment covered a much larger kinematic
range, up to $Q^2 = 2$ GeV$^2$ and $\epsilon$ values down to about
0.3. We expect similar systematic uncertainties related to data and
fiducial cuts and dead detector corrections as were determined for the
results presented here.  We expect to reduce the systematic
uncertainty for positron/electron luminosity differences to $\sim$1\%
for the full run by forming an additional double ratio of results for
the two different chicane polarities.  We also utilized a beam-profile
monitor at the downstream end of CLAS to verify that the ratio of
positron luminosity for a given chicane polarity to electron
luminosity in the opposite chicane polarity was flat to within
$\sim$1\% as a function of lepton energy. Analysis of these data is
underway and we expect final results soon.

Two other experiments are measuring $R$ to determine the TPE effect
using electron and positron beams at internal storage rings.  The
Novosibirsk group \cite{VEPP-3,nikolenko10,gramolin12} measured $R$ at
six different kinematic points and in 2012, the OLYMPUS Collaboration
\cite{olympus} took data at a single lepton beam energy for $Q^2<2.5$
GeV$^2$.  These experiments have very different systematic
uncertainties and kinematic coverages from the CLAS experiment.
 
These experiments will provide information that is vital to our
understanding of the electron-scattering process as well as our
understanding of the proton structure.  We have heard the common
statement that ``the electromagnetic probe is well understood.''
However, the discrepancy between Rosenbluth and polarization
measurements of the form-factor ratio indicates otherwise.  Indeed, if
we don't understand elastic electron scattering at high precision or
when higher order contributions become significant, then similar
measurements will be in doubt.  There are important
implications for many of the nuclear physics quantities being studied
ranging from high-precision quasi-elastic experiments to strangeness
and parity violation experiments.

\begin{acknowledgments}

We acknowledge the efforts of the staff of the Accelerator and Physics
Divisions at Jefferson Lab that made this experiment possible.  We are
especially grateful to the Hall B staff members who tirelessly
reconfigured the beamline and stacked (and restacked) shielding blocks.  
Thanks also to Dave Kashy who made the crucial suggestion of narrowing the 
post-chicane collimator. This work was supported by the U.S. Department 
of Energy and National Science Foundation, the Israel Science Foundation, 
the US-Israeli Bi-National Science Foundation, the Chilean Comisi\'on Nacional 
de Investigaci\'on Cient\'ifica y
Tecnol\'ogica (CONICYT) grants FB0821, ACT-119, 1120953, 11121448, and
791100017, the French Centre National de la Recherche Scientifique and
Commissariat a l'Energie Atomique, the French-American Cultural
Exchange (FACE), the Italian Istituto Nazionale di Fisica Nucleare,
the National Research Foundation of Korea, and the United Kingdom's
Science and Technology Facilities Council (STFC). The Jefferson
Science Associates (JSA) operates the Thomas Jefferson National
Accelerator Facility for the United States Department of Energy under
contract DE-AC05-84ER40150.
\end{acknowledgments}

\bibliography{TPETR}

\end{document}